# Islanding Strategy for Smart Grids Oriented to Resilience Enhancement and Its Power Supply Range Optimization

Yanhong Luo[1], Wenchao Meng[2], Xi Zhu[3], Andreas Elombo[4], Hu Rong[5], Bing Xie[6] and Tianwen Zhang[7]

[1] School of Information Science and Engineering, Northeastern University, Shenyang 110819, China; 939803797@qq.com(W.L.); 20225381@stu.neu.edu.cn(J.Z.); luoyanhong@ise.neu.edu.cn(Y.L.)
[2] State Grid Yingkou Power Supply Company, Yingkou, Liaoning, China; hz3_yk@ln.sgcc.com.cn
\* Correspondence: luoyanhong@ise.neu.edu.cn; Tel.: 86-13898195450

**Abstract:** With the increasing prevalence of distributed generators, islanded operation based on distributed generation is considered a vital means to enhance the reliability and resilience of smart grids. This paper investigates the main factors in islanding partition of smart grids and establishes a mathematical model for islanding division. A method to determine the maximum power supply range of distributed energy resources (DERs) based on the reachability matrix and power circle algorithm is proposed to improve computational efficiency. A dynamic programming method based on breadth-first search (BFS) is used to solve the islanding partition scheme, and a region correction method is applied to modify the maximum power supply area by considering controllable loads and prioritizing critical load restoration, thereby enhancing system resilience. Finally, simulation results verify the effectiveness of the proposed algorithm in improving smart grid resilience.

## 1. Introduction

After a fault occurs in a smart grid, all downstream users are affected by power disruption. To ensure continuous power supply to critical users, once the fault is identified and isolated, using distributed generators (DGs) to serve critical loads enables smart grids to operate in islanded mode [1,2], effectively enhancing system reliability and resilience [3,4].

Reference [5] proposes a rapid fault isolation and service restoration method combining active islanding effects with RCS, improving the resilience and quick recovery capability of smart grids. Reference [6] establishes a comprehensive optimization model for post-outage balance restoration in smart grids under extreme events, considering topological flexibility to form dynamic islands via reconfiguration and reducing outage time. Reference [7] points out that the islanding partition problem is essentially a 0-1 integer programming problem. Reference [8] verifies that fault recovery outcomes are influenced by DER capacity and installation locations. Reference [9] proposes an islanding partition strategy based on BFS, considering DER types and reserve capacity to ensure stable islanded operation and enhance resilience. Reference [10] uses the shortest path method to incorporate critical loads into islanded areas, ensuring their power supply. Reference [11] presents an islanding partition method for smart grids based on chance-constrained programming to overcome fluctuations from reduced power supply. Reference [12] establishes a time-series islanding model for smart grids based on soft switching, reducing fault duration during islanding via the operational flexibility of SOP. Reference [13] uses Continuous Operating Time (COT) as the objective function to evaluate island availability for continuous service restoration, while Reference [14] proposes a service recovery method leveraging V2G characteristics of electric vehicles (EVs) to enhance smart grid resilience.

This paper proposes a method to determine the maximum power supply range of DERs based on the reachability matrix and power circle algorithm, aiming to compress the solution space and improve efficiency. Additionally, a correction method for the maximum supply area is introduced by prioritizing critical loads and utilizing controllable loads, thereby increasing load recovery and enhancing smart grid resilience.

## 2. Modeling of Islanding Partition in Smart Grids

In scenarios with multiple faults, the islanding partition of smart grids significantly impacts load power restoration and system resilience enhancement. Loads in power systems have different importance levels due to varying impacts of power outages. After major faults, smart grids often cannot supply all loads, necessitating islanding partition to prioritize critical load restoration and enhance resilience. Key considerations for islanding partition include: safety of islanded areas, load restoration priority, algorithm efficiency, DER output uncertainty, and rational use of controllable loads—all directly related to resilience improvement.

If controllable loads exist, they should be used to reduce or interrupt non-critical loads for restoring more critical loads, a key strategy for resilience enhancement. The objective function for islanding partition is formulated to maximize critical load restoration, as shown in:

$$\max f = \sum_{i=1}^{n} x_i \omega_i p_i \tag{1}$$

Where, $f$ is the objective function, which mainly aims to maximize the restoration of power supply to critical loads. $x_i$ represents whether the power supply to node $i$ is restored; it takes the value of 1 if the power is restored, and 0 if it is not. $\omega_i$ denotes the load weight of node $i$, where loads are classified into primary, secondary, and tertiary categories, with corresponding weights of 100, 10, and 1, respectively. $p_i$ represents the load size at node $i$.

The constraints for the islanding division mainly consider the following aspects:

(1) DG capacity constraint

$$P_{DG,j} - \sum_{i=1}^{n}(x_j * p_i) \geq 0 \tag{2}$$

Where, $P_{DG,j}$ represents the output of distributed generation in the $j$ island, $n$ is the total number of nodes in the distribution network, and $x_j$ is a binary variable. If the load node is within the $j$ island, $x_j = 1$; otherwise, $x_j = 0$.

(2) Node voltage constraint

$$U_{\min} \leq U_i \leq U_{\max} \tag{3}$$

Where, $U_{\min}$ and $U_{\max}$ represent the lower and upper voltage limits for a node, respectively, and $U_i$ denotes the voltage at node $i$. The node voltage within the island should satisfy the safety constraints, ensuring that the voltage stays within the specified minimum and maximum limits.

(3) Branch current constraint

$$I_i < I_N \tag{4}$$

Where, $I_N$ represents the rated current of the line, and $I_i$ denotes the current on line $i$.

## 3. Calculation of Maximum Power Supply Area for Distributed Energy Resources Based on Reachability Matrix and Power Circle Algorithm

*This section presents a method to determine the maximum power supply range of Distributed Energy Resources using the reachability matrix and power circle algorithm, a core technique for resilience enhancement in smart grids.*

*3.1. Determination of Reachable Areas for Distributed Energy Resources Based on Reachability Matrix*

Based on the network $G(V,E)$, the adjacency matrix of the network $A$ can be computed, with the elements expressed as:

$$a_{ij} = \begin{cases} 1 & from\ node\ i\ to\ node\ j \\ 0 & others \end{cases} \quad (5)$$

Where, $a_{ij}$ represents the element in the $i$ row and $j$ column of the adjacency matrix. If there is an edge between node $i$ and node $j$ in the network, then $a_{ij}$; otherwise, $a_{ij}$. If the network is undirected, then $a_{ij}=a_{ji}$, $A = A^T$.

The reachability matrix can be obtained by calculating the powers of the adjacency matrix and performing a logical OR operation, as shown in:

$$P = \prod_{i=1}^{n} A^i \quad (6)$$

Where, $P$ represents the reachability matrix, while $\prod$ denotes the logical OR operation for matrices. $n$ represents the total number of nodes in the distribution network. The elements in $P$ are also 0-1 variables: if $p_{ij}=1$, it indicates a path exists between nodes $i$ and $j$; if $p_{ij}=0$, it indicates no path exists between nodes $i$ and $j$, meaning nodes $i$ and $j$ belong to two separate subnetworks.

*3.2. Determination of Maximum Power Supply Range for Distributed Energy Resources Based on Power Circle Algorithm*

When the distribution network operates in islanded mode, fluctuations in the output of distributed energy resources may lead to failure in power restoration and even cause more severe faults. Therefore, it is essential to consider the uncertainty of distributed energy resources output during island partitioning. The fluctuations in distributed energy resources output at a given moment often follow a normal distribution, which can be expressed as:

$$f(x) = \frac{1}{\sqrt{2\pi}\sigma} e^{\frac{(x-\mu)^2}{2\sigma^2}} \quad (7)$$

Where, $\mu$ represents the expected value of the distributed energy resource output, and $\sigma$ denotes the standard deviation of the distributed energy resource output.

To effectively enhance the reliability of island partitioning and ensure stable power supply despite output fluctuations, the actual stable output of distributed energy resources should be considered. This is calculated by subtracting the power reduction due to fluctuations from the predicted output power of the distributed energy resources, which can be expressed as:

$$P_{DG} = P_{DG.pre} - \sigma_{DG} \quad (8)$$

Where, $P_{DG}$ represents the actual stable output of the distributed energy resource, $P_{DG.pre}$ denotes the predicted output of the distributed energy resource, and $\sigma_{DG}$ represents the standard deviation of the output, serving as the reduction amount in the distributed energy resource's output.

To accurately and quickly determine the maximum supply range of an islanded distribution network, this study adopts the "power circle" method, with the distributed energy resource as the center, to identify the distributed energy resources' maximum supply area. In distribution networks, multiple distributed energy resources often exist, and the maximum supply ranges obtained from the power circle search method centered on different distributed energy resources may overlap. This overlap indicates surplus output from the distributed energy resources. Therefore, when the maximum supply ranges of two or more distributed energy resources intersect, the surplus power of these distributed energy resources can be used to further expand the supply range.

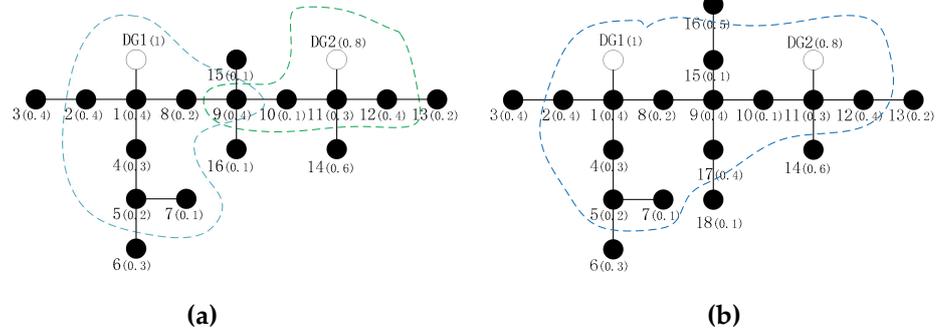

**Figure 1.** (a)Example diagram for the case where the power circles are crossed. (b) Map of power supply area after merging based on power circles

As shown in Figure 1.(a), there is an overlap between the power circles of DG1 and DG2. Therefore, when calculating the supply range, these two power circles can be merged. The remaining power can then be used to re-calculate the power circle, using the overlap area as the root node. This calculation considers the load demand within the overlapping area and the total remaining power of each distributed energy source after supplying this area, allowing for the identification of additional areas that can be supplied by the merged power circles.

As shown in Figure 1.(b), based on the overlapping maximum supply ranges in Figure1.(a), the load demand within the overlapping area is calculated to be 0.4, and after supplying this area, DG1 and DG2 have no remaining power. Thus, using node 9 as the root node with a supply capacity of 0.4, a power circle search is conducted. The additional supply area is then combined with the original supply area calculated for the power circle, resulting in the maximum supply range of the merged distributed energy resources. The final result is illustrated in Figure 1.(b).

## 4. Islanding Partition Method for Smart Grids and Resilience Enhancement Strategy

The islanding partition method proposed in this paper rounds the power of both power sources and loads while incorporating resilience enhancement strategies. For the power output of source nodes, downward rounding is used, whereas upward rounding is applied to load power to avoid voltage violations, ensuring the operational stability of smart grids, which serves as the foundation for resilience enhancement. However, due to the presence of controllable loads in smart grids, rational utilization of these controllable loads for power reduction can increase the recovery of critical loads, directly enhancing system resilience. The controllable loads are decomposed as shown in Equation (9):

$$P_{L,i}=P_{L,i,c}+P_{L,i,n} \qquad (9)$$

Where, $P_{L,i}$ represents the load at node $i$; $P_{L,i,c}$ represents the reducible load at node $i$; and $P_{L,i,n}$ represents the non-reducible load at node $i$. If node $i$ is an uncontrollable load, then $P_{L,i,c}=0$, and $P_{L,i}=P_{L,i,n}$; if node $i$ is a fully controllable load, then $P_{L,i,n}=0$, and $P_{L,i}=P_{L,i,c}$.

Furthermore, in the hierarchical partition based on breadth-first search, controllable loads are decomposed into two parts, which is a key strategy for resilience enhancement. Suppose nodes 2 and 3 are controllable load nodes. When considering power supply to the critical load node 4, nodes 2 and 3 can be decomposed such that the controllable loads of nodes and represent the reducible or interruptible load amounts, while the non-interruptible portions of the controllable loads at nodes 2 and 3 are also identified. Through this decomposition, elastic scheduling of load resources is achieved, enhancing the recovery capability of smart grids.

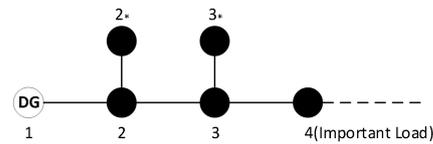

**Figure 2.** Schematic diagram of controllable load decomposition.

For the decomposed maximum power supply area of distributed energy, this paper employs the breadth-first search method to perform hierarchical processing of the maximum supply area of distributed power sources, providing structural optimization support for resilience enhancement. As shown in Figure 3, network nodes are layered based on the number of nodes traversed when connected to the root node.

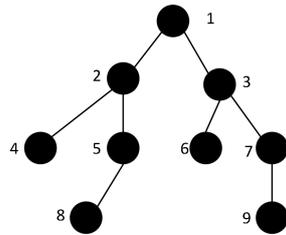

**Figure 3.** Example diagram of hierarchical processing based on breadth-first search.

After hierarchical calculation of the network, the concept of dynamic programming is introduced into the islanding partition solution method to achieve resilience enhancement through dynamic adjustment. The solution process backtracks from the lowest layer until reaching the layer where the root node is located, completing the solution, which reflects the dynamic adaptability of smart grids to faults.

The primary goal of islanding partition in smart grids is to maximize the power supply recovery of critical loads, which is the core objective of enhancing system resilience. Therefore, all critical loads within the maximum power supply area of distributed energy are prioritized for restoration. Loads in this paper are classified into three types: primary loads, secondary loads, and tertiary loads. When verifying critical loads, the first priority is to verify whether the output power of distributed energy can meet the capacity constraint for restoring all primary loads, ensuring power supply to critical loads and enhancing the survival resilience of smart grids. If the capacity constraint is satisfied, the restoration of all secondary loads within the supply range is considered next. If all secondary loads can be restored, it is determined that all secondary loads and the loads along the path connecting the secondary loads to the power source node can be restored, achieving elastic expansion of the power supply range. If the power constraint is not satisfied, the new power supply load restoration in the current verification area is compared with that in the previous load verification area to evaluate the power supply situation, and the system's resilient adaptability is enhanced through dynamic adjustment strategies.

**5. Case analysis**

The IEEE 69-bus system with multiple distributed energy resources is used to verify the effectiveness of the proposed islanding partition method and its resilience enhancement for smart grids. The IEEE 69-bus system has a primary-side node network voltage of 12.66 kV, consisting of 69 nodes and 68 branches, with node 1 connected to the superior grid. The total load in the network is 3802.19 kW + j2694.69 kVar. In the IEEE-69 bus system, six distributed power sources capable of independent power supply are configured. The specific nodes and rated capacities are shown in Table 1. Load priorities are listed in Table 2, and the load type of each node is presented in Table 3.

**Table 1.** Distributed energy configuration parameters

| The number of DG | Node | Rated capacity/kW |
|---|---|---|
| DG1 | 5 | 250 |
| DG2 | 19 | 400 |
| DG3 | 32 | 40 |
| DG4 | 36 | 50 |
| DG5 | 52 | 1300 |
| DG6 | 65 | 100 |

**Table 2.** IEEE 69 node significance levels for each load

| Load level | Load weighting | Node |
|---|---|---|
| Primary load | 100 | 6,9,12,18,27,35,37,42,51,57,62 |
| Secondary load | 10 | Others |
| Tertiary load | 1 | 7,10,11,13,16,22,28,43,48,59,60,63 |

**Table 3.** IEEE 69 nodes for each load type

| Load type | Node |
|---|---|
| 100%controllable | 8,11,13,21,26,28,34,35,37,39,40,43, 48,54,55,56,66,67,68,69 |
| uncontrollable | Others |

Assume that a fault occurs on the line between node 3 and node 4 in the distribution network.

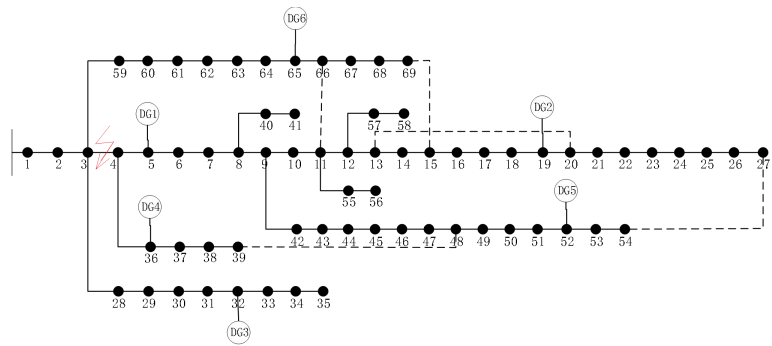

**Figure 4.** IEEE 69 node single point of failure example diagrams

First, the reachable areas of distributed power sources in the smart grid are determined based on the reachability matrix. The reachable areas are shown in Figure 5.

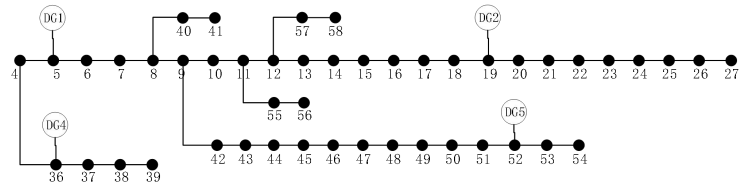

**Figure 5.** Distributed energy accessible areas

According to the reachability matrix, due to the extreme fault, the line fault divides the smart grid into two parts. Figure 5 shows the reachable areas of distributed power sources disconnected from the main grid, including DG1, DG2, DG4, and DG5. Meanwhile, DG3 and DG6 remain connected to the main grid and do not require islanding partition, demonstrating the resilience maintenance capability of smart grids under partial faults.

The maximum power supply area of each distributed energy resource determined by the power circle method is as follows: the maximum supply area of DG1 includes nodes 4, 5, 6, 7, 8, 9, 10, 36, 37, 40, and 41; the maximum supply area of DG2 covers nodes 9, 10, 11, 12, 13, 14, 15, 16, 17, 18, 19, 20, 21, 22, 23, 24, 25, 26, 27, 55, 56, 57, and 58; the maximum supply area of DG4 spans nodes 4, 5, 6, 7, 36, and 37; the maximum supply area of DG5 includes nodes 44, 45, 46, 47, 48, 49, 50, 51, 52, 53, and 54.

Next, for the maximum supply areas of DG1, DG4, and DG5, which contain relatively few nodes, the actual islanding partition method is directly calculated using the breadth-first dynamic programming algorithm. For the maximum supply area of DG2, which contains more nodes, the supply area is modified using the method proposed in this paper that considers critical load priority, which is one of the core strategies for enhancing smart grid resilience. This method includes confirming the restoration of primary loads and nodes along the connection path to the power source, ensuring the priority restoration of critical loads and enhancing system resilience. The supply area confirmed based on primary load restoration is shown in Figure 6.

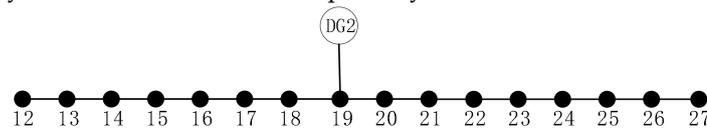

**Figure 6.** Confirmation of restored power supply areas based on power supply area correction

After narrowing the supply area, the breadth-first dynamic programming algorithm is used to calculate the actual islanding partition scheme of distributed energy resources. The islanding partition scheme is shown in Figure 7. The specific load recovery results are presented in Table 4.

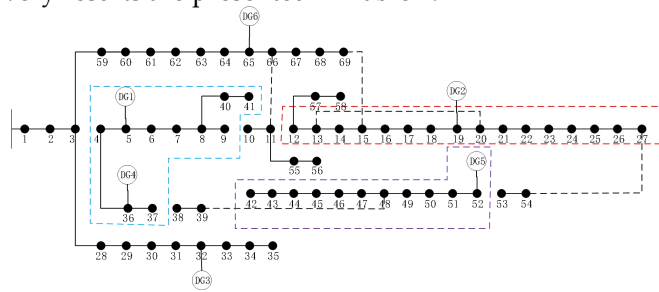

**Figure 7.** Island division program

**Table 4.** Island division results

| Islanding area | The number of DG | Power restoration load node |
|---|---|---|
| 1 | DG1、DG4 | 4-9、36-37、40-41 |
| 2 | DG2 | 12-20、22-27 |
| 3 | DG5 | 42、44-47、49-52 |

The islanding partition method proposed in this paper restores loads at all levels, as shown in Table 5. It restores 366.95 kW of primary loads, 1498.1 kW of secondary loads, and 99.2 kW of tertiary loads, with a total load restoration of up to 1964.25 kW. In terms of the recovery rate, the primary load recovery rate reaches 86%, significantly enhancing the critical service maintenance capability of smart grids after faults and demonstrating the effectiveness of resilience enhancement.

**Table 5.** Amount of load restoration at all levels

| Load level | Total load/kW | Load restoration/kW | Restoration ratio /% |
|---|---|---|---|
| Primary load | 424.95 | 366.95 | 86 |
| Secondary load | 2876.64 | 1498.1 | 52 |
| Tertiary load | 500.6 | 99.2 | 20 |

## 6. Conclusions

Aiming at the resilience enhancement needs of smart grids, this paper calculates the maximum power supply range of distributed energy using methods based on the reachability matrix and power circle, solving the problem of excessively large solution space in islanding partition and improving computational efficiency, thus providing technical support for resilience enhancement. Then, the maximum power supply range is adjusted by considering correction methods based on controllable loads and priority restoration of critical loads, achieving elastic scheduling of power supply resources. For areas where power supply is uncertain, the breadth-first search dynamic programming algorithm is used to determine the optimal supply area, enhancing the dynamic adaptability of smart grids to faults. Finally, simulation analysis verifies the effectiveness of the proposed algorithm in enhancing smart grid resilience, providing important references for the reliability design and operation of smart grids.

**Author Contributions:** Conceptualization, Weiyan Liu and Zhe Hu; methodology, Weiyan Liu; software, Zhe Hu; validation, Weiyan Liu, Zhe Hu and Jiubo Zhang; formal analysis, Zhe Hu; investigation, Jiubo Zhang; resources, Weiyan Liu and Zhe Hu; data curation, Weiyan Liu and Zhe Hu; writing—original draft preparation, Weiyan Liu and Zhe Hu; writing—review and editing, Weiyan Liu, Zhe Hu and Jiubo Zhang; visualization, Jiubo Zhang; supervision, Yanhong Luo; project administration, Yanhong Luo; funding acquisition, Yanhong Luo. All authors have read and agreed to the published version of the manuscript.

**Funding:** This research was funded by the National Natural Science Foundation of China, grant number (62173082, U22B20115) and the National Key Research and Development Program of China, grant number (2022YFB4100802).

**Data Availability Statement:** The raw data supporting the conclusions of this article will be made available by the authors on request.

**Acknowledgments:** The authors would like to thank their research institutions and supervisors as well as the reviewers who provided valuable suggestions to this paper.

**Conflicts of Interest:** The authors declare no conflicts of interest.